\documentstyle[11pt,paspconf]{article}

\begin{document}

\title{Novae and BAL QSOs: The Aluminum Test}

\author{Gregory A. Shields}

\affil{Department of Astronomy,
The University of Texas, Austin, TX 78712}

\begin{abstract}
Novae have been proposed as the explanation of high reported abundances of
heavy elements 
in the gas producing the broad absorption lines
(BALs) of QSOs.  High
abundances of odd numbered elements, including aluminum, are predicted.
Available data contains
hints that the Al/Si ratio may be high both in the BAL gas and in the broad
emission line (BEL) gas.
\end{abstract}


\section{Abundances in BAL QSOs}

Broad absorption lines caused by rapidly (${\mathrel{\vcenter
     {\hbox{$<$}\nointerlineskip\hbox{$\sim$}}}}$30,000 km s$^{-1}$)
outflowing gas are seen in
the spectra of $\sim$10\% of radio quiet QSOs (Weymann et al. 1991).
Analysis of the derived
column densities of H$^0$ and various ions of the heavy elements have led to
reported abundances
of C, N, O, Si, and sometimes other elements, that are 1 to 2 orders of
magnitude greater than
solar (Turnshek et al. 1996, and references therein).  A startlingly high
abundance of
phosphorus, P/C $\approx$ 65 (P/C)$_\odot$, was reported by Junkkarinen et
al. (1995; see also
Junkkarinen et al. in these Proceedings).  The problems of scattered light and
incomplete geometrical
coverage of the continuum source by the BAL gas introduce serious
uncertainties (Arav in
these Proceedings).  Nevertheless, the extraordinary abundances reported
for the BAL gas seem
likely to survive, at least qualitatively, and the phosphorus anomaly in
particular points to an
exotic origin.

\section{Novae in AGN}

Shields (1996) has proposed that the BAL gas largely consists of debris of nova
explosions occurring
in the inner few light years of the QSO nucleus.  This is motivated by high
phosphorus
abundances in the ejecta of model novae (Politano et al. 1995) and by the
resemblance of C, N, O, and Si
abundances in observed nova shells to those in BAL QSOs.  Approximately
one nova per year is
required to produce the BAL gas if the BALs occur at a radius
of $\sim10^{18}$~cm, 
the minimum allowed by the fact that BALs sometimes absorb the
broad line emission. This rate of novae could 
occur in a nuclear star cluster of mass
$\sim10^8$ M$_\odot$ in which single white dwarfs accrete the necessary
layer of hydrogen by
means of repeated orbital passages through an accretion disk around a
supermassive black hole.

\section{Aluminum in QSOs}

 A prediction of nova models is
that the odd numbered elements will have enhanced abundances, relative to
neighboring even
elements, in comparison with normal cosmic abundances.  For example,
Politano et al. (1995)
predict P/C values of 50 and 300 times solar for
explosions on white
dwarfs of 1.25 and 1.35 M$_\odot$, respectively.  High Al is observed
in nova debris (Andre\"a et al. 1994).  Al~III
$\lambda1857$ is seen both in BAL and BEL spectra.  This offers a potential
test of novae as a
source of BAL gas and as a contribution to the BEL gas.

\subsection{Broad Absorption Lines}

BALs typically involve fairly high stages of ionization, e.g., C IV, Si~IV,
N V, and O VI.
Photoionization models (e.g., Turnshek et al. 1996) 
typically involve gas that is optically thin in the
Lyman continuum of H
I, $\tau_{\rm H}<<1$, consistent with the measured column densities of H I.
On the other hand, the low ionization BAL
QSOs (``Lo-BALs" or
``Mg II BAL QSOs") show a variety of low ionization lines, often including
Al~III~$\lambda$1857
(Wemann et al. 1991; Voit et al. 1993).  
From the spectra of Wemann et
al. (1991), I estimate $\tau_{\rm Al~III}/\tau_{\rm Si~IV}\approx$ 
0.30, 0.30, and 0.36 for QSOs 1011+091, 
1231+1325, and 1331--0108 respectively.  
These values correspond to column density ratios
Al$^{+2}$/Si$^{+3}$ =
0.22, 0.22, and 0.27.
Figures 2a,b of Voit et
al. (1993) give BAL optical depths for models of photoionized clouds with 
the Mathews and Ferland (1987, ``MF") ionizing continuum 
 and ionization paramters log U = -1.5 and - 1.0, where
${\rm U}\equiv\phi_{UV}/{\rm Nc}$, 
$\phi_{UV}$ is the incident
flux of ionizing photons (cm$^{-2}$ s$^{-1}$), and N is the number density
of atoms.  
For both values of U,
these models give $\tau_{\rm Al~III}/\tau_{\rm Si~IV}\approx0.05$ 
for solar Al/Si
and an ionization structure extending well into the He$^+$ zone
(consistent with the
presence of low ionization lines).  This suggests that the 3 QSOs may have
Al/Si $\approx$
6(Al/Si)$_\odot$.

Unfortunately, this simple approach suffers the uncertainty noted above
concerning scattered
light affecting the minimum trough intensity.  In order to minimize this
concern, one may look
for cases in which the strongest BALs reach quite low minimum intensities
relative to the
continuum, I$_{\rm min}$/I$_{\rm cont}$, but both the Si~IV and Al~III
lines have substantially
shallower BALs.  The scattered light would then give only a modest
contribution to I$_{\rm min}$
in the Si~IV and Al~III lines, and the derived optical depths would be
meaningful in the context
of the gross abundance anomalies of interest here.  An interesting
candidate is UM 232 (Q0019+011), a
$z_{e}\approx2.12$ BAL QSO for which Barlow et al. (1989) observed Si~IV
and Al~III absorption
considerably shallower than C IV.  Moreover, the equivalent widths of the
BALs descreased with
increasing continuum luminosity, $L$, in a way suggestive of a changing
photoionization
equilibrium.  Adopting this interpretation, I have computed a simple set of
photoionization
models using the CLOUDY program (Ferland 1996).  
The models had
$\tau_{\rm H}<<1$, solar abundances, a broken power-law continuum,
and a range of ionization parameters.
(The abundances have relatively little effect on the ionization for small
$\tau_{\rm H}$.)  From the
observed fact that the C IV equivalent width decreased with increasing $L$,
but less so than Al~
III and Si~IV, one may estimate $U\approx10^{-1.9}$ from the fractional
abundances
$X(\rm A^{+i})\equiv N(\rm A^{+i})/N(\rm A)$ in the models.  
This in turn implies an
ionization correction,
$X({\rm Si}^{+3})/X({\rm Al}^{+2})\approx5$.  From the column density ratios
${\rm Al}^{+2}/{\rm Si}^{+3}\equiv N({\rm Al}^{+2})/N({\rm Si}^{+3})=0.18$
given by Barlow et al.
(1989), we find
${\rm Al/Si}\approx0.9$, ten times the solar value ${\rm Al/Si}=0.09$
(Anders and Grevesse 1989).

This result has various uncertainties.  Subsequent observations of UM232,
and observations of
other variable BAL QSOs, do not in general show a clean correlation of BAL depth
with luminosity (Barlow et al. 1992; Barlow
1997).  Low ionization material has a lower value of $X({\rm
Si}^{+3})/X({\rm Al}^{+2})$, and a
model with a mix of high U and low U material might give a lower derived
Al/Si.  Nevertheless, the
order-of-magnitude Al/Si excess found here by a straightforward analysis
provides an intriguing
hint that Al as well as P may be enhanced in some BAL QSOs.

\subsection{Broad Emission Lines}

A recently popular view of the abundances in the broad emission-line
region (BLR) holds that rapid star formation
quickly produces a
metallicity $Z\approx5$ to 10 Z$_\odot$, following a fairly normal chemical
evolution
scenario (Hamann and Ferland 1993).  This view largely rests on the strong
observed nitrogen
lines together with the expected increase in N/C and N/O with increasing Z
because of secondary
production of N.  (Note that high value of N/C and N/O are also found in
nova debris.)
However, the inferred dimensions of the BLR
are only slightly less
than those assumed here ($\sim10^{18}$ cm) for the BAL region, and the BLR
involves quite
modest amounts of gas ($\sim$10 to 100 M$_\odot$).  If only a fraction of
the BLR gas 
comes from the same source as the BAL gas, elevated
abundances of P, Al,
and other odd numbered elements could result.

The QSO literature contains scattered remarks to the effect that the Al~III
$\lambda$1857
emission feature is much stronger than expected for solar abundances
(Uomoto 1984; Boyd and
Ferland 1987).  Weymann et al. (1991) find that Al~III is especially strong
in BAL QSOs,
although confusion with Fe emission is an issue.  Could these observations find a
straightforward
explanation in terms of enhanced abundances of Al?

With Fred Hamann of UCSD, I have begun to look at the possibility that gas
with abundances
resembling those reported for the BAL gas could 
contribute substantially to the
BLR.  Here I focus on the specific question of whether Al is enhanced
compared with its even
numbered neighbor, Si.  We have computed a simple grid of BLR
photoionization models using
CLOUDY.  The models  have a
slab geometry with a gas density $\rm N_H = 10^{10.5}~cm^{-3}$ and
log U = -3.0, -2.0, -1.0, and 0.0.  
The ionizing continuum is that of Mathews and Ferland (1987).
Only the  model with log U = 0.0 remained ionized to the stopping column density
of $10^{23.5}~\rm cm^{-3}$.
Laor et al. (1995)
quote Al~III intensities for a number of QSOs.  For these objects, average
values are Al~III/(Si~III + Si~IV)
 $\approx$ 0.30 and Si~III]/Si~IV $\approx$ 0.5.  The solar abundance
models reproduce the
Si~III]/Si~IV ratio for log U slightly greater than -1.0.  The models show
a substantial
increase in Al~III/Si~III] and decrease in Al~III/Si~IV with increasing U,
but Al~III/(Si~III] + Si~IV)
 is more stable, having values 0.08, 0.07, 0.06, and 0.03 for log U = -3,
-2, -1, and 0,
respectively.  From the model value for log U = -1 and the 
above value of Al~III/(Si~III]
 + Si~IV), we find [Al/Si] $\equiv \rm log_{10}[(Al/Si)/(Al/Si)_\odot]
\approx$ +0.7, assuming the the line ratio
scales linearly with the abundance ratio.

In order to test the robustness of this result, we constructed a composite
model with equal
amounts of ionizing continuum intercepted by solar abundance clouds with
log U = -2, -1, and
0.  This is motivated by the ``locally optimally emitting clouds" (LOC)
picture proposed by
Baldwin et al. (1995).  This model gave Si~III]/Si~IV = 1.0, higher than
observed; but it had
Al~III/(Si~III] + Si~IV) identical to the log U = -1 model, implying the
same Al/Si.

Analysis of a set of BLR models with abundances based on measured BAL 
abundances is in progress, with the goal of assessing the overall
compatibility with observed broad emission-line intensities and the
dependence of the derived Al/Si ratio on overall metallicity.
Another issue is the possibility of very high densities in QSO
emission-line regions.
Baldwin et al. (1996) developed a 3 component model to explain the broad
emission-line
intensities and profiles of 7 QSOs.  Their component ``A" has
$N_H\sim10^{12.5}$ cm$^{-3}$ and
emits strong Al~III.  This component's parameters have largely
been set to explain
the observed Al~III intensity.
A high aluminum abundance 
may offer an explanation of the strong observed Al~III emission
without recourse to such complex geometries.  However, the recent
observations of the Al~III and Mg~II doublet ratios in I Zw 1 by
Laor et al. (1997) show substantial thermalization
(see also Baldwin et al. 1996).  This may indicate
a high density in the Al~III emitting region; and in any case,
thermalization will affect the line ratios.

In summary, there are indications that Al/Si may be several times the solar
value in the BLR of
some QSOs.  However, quantitative results are difficult to obtain because
of the possible effect of
overall high metallicities on the ionization structure and the possibility
of high densities and thermalization.

\section{Conclusions}

The original impetus to consider novae as a source of enriched gas in QSOs
was the high P/C
ratio reported by Junkkarinen et al. (1995).  The high reported abundances
of the more common
heavy elements in the BAL gas, while still uncertain, resemble those of
novae, and accretion
onto single white dwarfs in the inner light year of 
the QSO nucleus provides a possible
mechanism for frequent nova eruptions.  The high predicted abundances of
odd numbered elements
in nova debris can be tested by observations of Al~III $\lambda$1857
absorption and emission.
The favorable case  of UM 232 gives [Al/Si] $\approx$ +1.0.  The strong
emission observed in Al~
III in many QSOs suggests [Al/Si] $\approx$ +0.7 by a
straightforward analysis, but
various complications could reduce this value.  
Nonetheless, these results hint at the
possibility of high
aluminum abundances both in the absorbing and emitting gas of QSOs.
Further observational and
theoretical work is needed to confirm the Al/Si results as well as to
explore the prevalence of
high phosphorus and to test for enhancements of other 
odd numbered elements.  Such
confirmation would
lend support to the idea that nova explosions (or some other high
temperature,
nonequilibrium hydrogen burning process) are an important source of gas in QSOs.

\acknowledgments
This material is based in part upon work supported by
the Texas Advanced Research Program under Grant No.
003658-015.

\begin{question}{Dr.\ Burbidge}
In the original nucleosynthesis work, it was slow neutron capture
(the s-process for elements in the Ne--Si range) that contributed
most to the odd-even element differences.  We always had a problem
with phosphorus, that is, accounting for its abundance in the
solar system abundances.
\end{question}
\begin{answer}{Dr.\ Shields}
The solar system phosphorus abundance can be explained by 
production in massive stars, mainly neon burning before the
star explodes (Woosley and  Weaver 1995; Timmes et al. 1995).
There are still odd-even abundance differences in the theoretical
nova nucleosynthesis results (e.g., Politano et al. 1995).
However, nucleosynthesis in nova explosions
occurs under nonequilibrium conditions, and the abundances depend
largely on cross sections.  This gives less of an odd-even effect than for
equilibrium nucleosynthsis, in which the odd-even effect reflects
differences in binding energy that have an exponential effect on
abundances.
\end{answer}
 
\begin{question}{Dr.\ Foltz}
Do you have enough cosmic time to produce the white dwarfs?
\end{question}
\begin{answer}{Dr.\ Shields}
The Universe was several billion years old when the typical BAL QSO
emitted the light we see, so there was plenty of time for white dwarfs
if the galaxy formed at a redshift of z = 5 or 10.  Even if the
nuclear star cluster formed at the beginning of the observed QSO outburst,
and this lasts only $\sim 10^8$ yr, 
there could be a substantial population
of white dwarfs.
\end{answer}

\end{document}